\documentclass[sigconf]{acmart}
\AtBeginDocument{%
  }

\copyrightyear{2026}
\acmYear{2026}
\setcopyright{cc}
\setcctype{by-nc-nd}
\acmConference[UIST '26]{The 39th Annual ACM Symposium on User Interface Software and Technology}{November 02--05, 2026}{Detroit, MI, USA}
\acmBooktitle{The 39th Annual ACM Symposium on User Interface Software and Technology (UIST '26), November 02--05, 2026, Detroit, MI, USA}
\acmDOI{10.1145/3830398.3830713}
\acmISBN{979-8-4007-2856-3/2026/11}

\usepackage{tabularx}
\usepackage{booktabs}
\usepackage{soul}

\begin{document}

\title{Camera-Free Object Discovery and Interaction with \textit{GazeTags}}  
\author{Vimal Mollyn}
\affiliation{
 \institution{Carnegie Mellon University}
 \city{Pittsburgh}
 \state{PA}
 \country{USA}}
\email{vmollyn@cs.cmu.edu}

\author{Chris Harrison}
\affiliation{
 \institution{Carnegie Mellon University}
 \city{Pittsburgh}
 \state{PA}
 \country{USA}}
\email{chris.harrison@cs.cmu.edu}

\renewcommand{\shortauthors}{Mollyn and Harrison}

\begin{abstract}
We present GazeTags, a camera- and computer-vision-free approach for worn head-gaze object awareness. Our small electronic tags cost roughly \$2 each and can operate for more than a year on a coin cell battery. They work in concert with a user-worn device (smartglasses, hearables, etc.) that periodically emits a focused infrared signal aligned with the user’s head direction. Only GazeTags within that narrow head-oriented beam wake and respond, providing their identity and optional payloads such as object state or live sensor readings. By avoiding cameras and computer vision pipelines, our approach is very energy efficient, able to operate continuously even in battery-constrained wearables, and sidesteps non-trivial privacy issues. We describe our hardware design, characterize our system performance and operating tradeoffs, and demonstrate how GazeTags can enable practical gaze-aware environments. 
\end{abstract}



\begin{teaserfigure}
  \includegraphics[width=\textwidth]{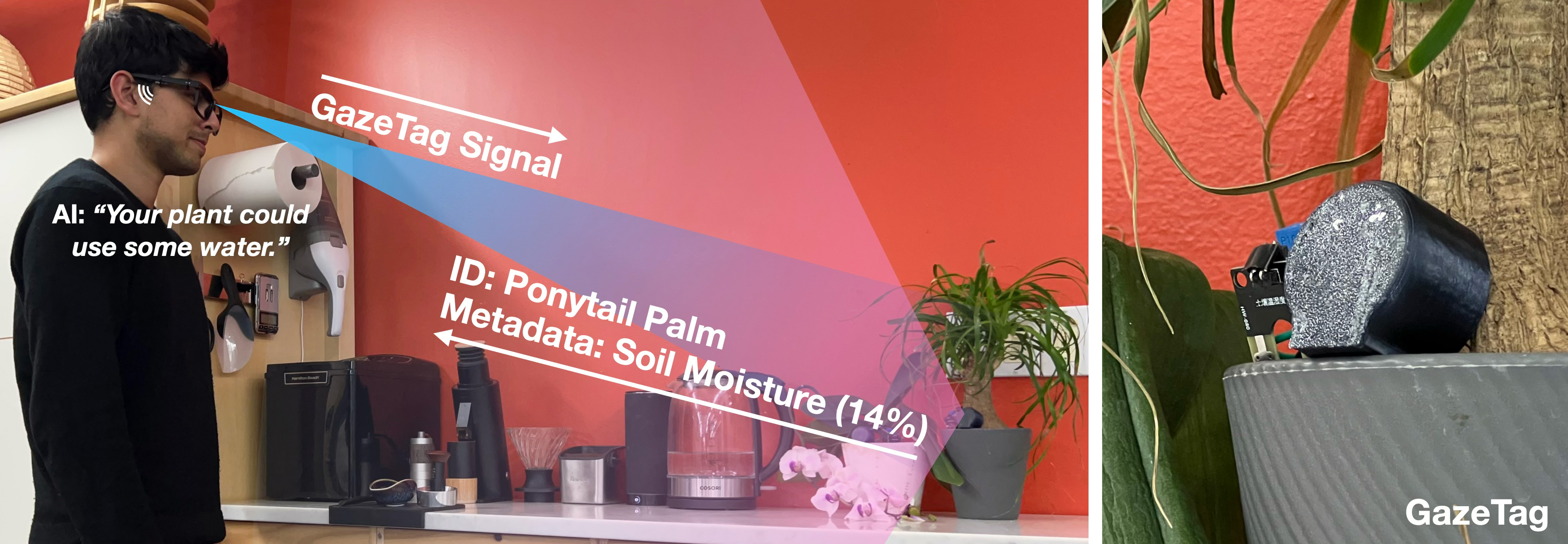}
  \vspace{-5mm}
  \caption{GazeTags enable head-oriented gaze-aware environments and interactions \textit{without cameras}. A narrow infrared signal emitted from a user's smart glasses (or earbuds) activates GazeTags the user is facing. In this example, a GazeTag attached to a potted plant reports back its object type (Ponytail Palm) and metadata (14\% soil moisture). Using this information, a passively-monitoring AI agent could \textit{proactively} prompt the user to water the plant. Or more conventionally, the user could also ask their AI agent if their plant needs water. At roughly \$2 per GazeTag, which runs for more than a year on a coin cell battery, environments could be densely instrumented with smart functionality.}
  \vspace{3
  mm}
  \Description{}
  \label{fig:teaser}
\end{teaserfigure}

\maketitle

\section{Introduction}
\begin{figure*}[t]
    \centering
    \includegraphics[width=\linewidth]{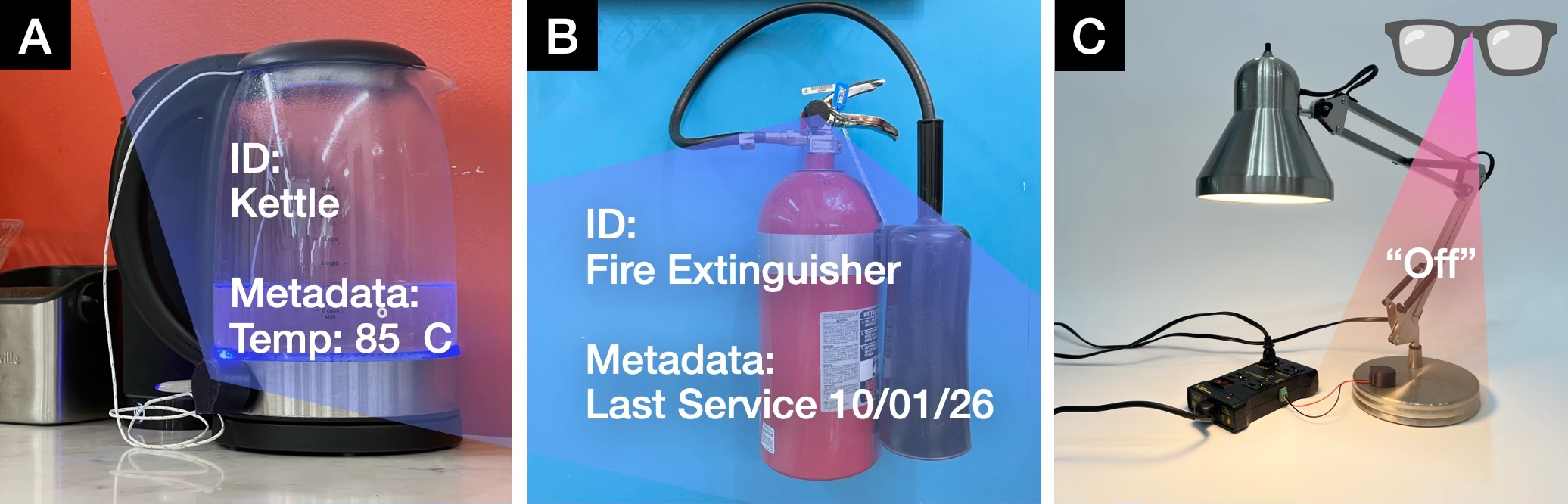}
    \caption{Example GazeTag uses. (A) A kettle equipped with a GazeTag and thermistor reports its current temperature. (B) A fire extinguisher provides its last serviced date to a health and safety inspector. (C) A user instructs a GazeTag equipped lamp to "turn off".}
    \label{fig:example-apps}
\end{figure*}

Smart glasses, AI pins, earbuds and other head-worn devices offer the promise of always-available and proactive assistance. To achieve their full potential, these systems must understand what in the environment is relevant to the user. Gaze is one of the strongest signals for this purpose, often revealing a user's intent or task. This can be fine-grained eye-vector gaze (what specific object are they looking at) or coarser-grained head-vector gaze (what part of a scene are they facing towards). Systems with this information can adapt their user interfaces and offer timely assistance, such as retrieving information about appliances, surfacing instructions for tools, or reacting to changes in attended objects. 

Today, however, gaze-aware wearables generally rely on cameras and computer vision. While powerful, this is especially problematic for small, battery-constrained wearable platforms. Cameras are relatively power-hungry sensors, and vision pipelines are also energy expensive, either when run locally or requiring transmission elsewhere for processing. As a result, continuous operation remains an outstanding challenge for compact devices such as glasses, earbuds, and pins. In practice, this most often means that vision-based sensing is either explicitly triggered by wearers for short periods, or launched periodically with a low duty cycle, limiting its usefulness for always-available contextual assistance.

A second major challenge of vision-powered gaze for wearables is privacy. Even when such systems are technically effective, their social acceptability can be limited by discomfort around image capture, uncertainty about what is being recorded, and concern over downstream analysis. This tension is particularly acute for wearable AI, where the ideal technical solution is \textit{persistent} vision sensing. This could include video capture in sensitive contexts like bathrooms, where AI assistance could be invaluable for some users. However, active cameras in these contexts are presently non-starters for many users and bystanders. 

In this paper, we explore a different path: achieving head-oriented gaze sensing \textit{without} cameras or computer vision, sidestepping many significant power and privacy issues. Our proposed solution is GazeTags, a low-cost tagging approach for instrumenting environments so that head-worn devices can discover and interact with attended objects. A user-worn device periodically emits a narrow field-of-view infrared signal that is naturally aligned with the user’s head direction. Only tags within that head-gaze-correlated infrared ``spotlight’’ wake and respond. In reply, tags announce both identity and dynamic payloads, such as object state, sensor readings, and contextual metadata. In this way, GazeTags can even move beyond computer-vision sensing, exposing non-visual states of objects, such as temperature, moisture, and power draw. 

Of course, in order to be practically deployed across e.g., scores of objects in a home, GazeTags must be \textit{ultra-low-cost}, \textit{small}, and \textit{maintenance-free} over long periods. Our implementation achieves these three properties. GazeTags cost around \$2 USD each, including batteries, and could fall below \$1 in high volumes. This makes dense instrumentation of everyday environments feasible (e.g., a home kitchen with 15 appliances would cost around \$30 to augment). Second, GazeTags are small pucks, measuring 30~mm in diameter and 18~mm thick. They can be discreetly attached to objects of interest with magnets or tape. Third, GazeTags run for more than one year on a coin cell battery; when depleted, the battery can be replaced in seconds. 

At a high level, our approach offers several advantages. Foremost, GazeTags do not require cameras, image processing, or computer vision models, and they are far better matched to the energy and compute budgets of small wearable devices. They also avoid many of the privacy concerns associated with always-on visual sensing, since no imagery needs to be captured or analyzed. Yet, GazeTags preserve a form of gaze-aware object discovery and interaction that is highly useful for wearable interfaces. 

In the next section we highlight potential GazeTag capabilities through a set of example scenarios. We then review related work, focusing on other tag techniques and gaze sensing approaches. We detail our implementation, followed by a multi-part evaluation. Overall, results show our system works up to ranges of 8~m and with many GazeTags operating simultaneously. 

\section{Example GazeTag Uses}\label{sec:example-apps}
Although simple in design, GazeTags enable a variety of powerful interactions. To help illustrate this breadth of capability, we created a series of example uses (Figures~\ref{fig:teaser} and \ref{fig:example-apps}). These are fully functional demos; please also see the Video Figure. 

\subsection{Object-Aware Assistants}
We know from prior work that patterns of activity are powerful inputs to context-aware interfaces and AI assistants \cite{tapia2004activity,vankasteren2008accurate}. Thus, a simple yet powerful use case for GazeTags is to provide a real-time and continuous stream of what objects the user is facing. This could help an AI assistant stay informed about a user's current indoor location, task, and potentially intent: The user just entered the kitchen, just looked towards their coffee machine, etc.

\subsection{Objects with Sensors \& State}
Beyond transmitting object identity, GazeTags can also interface with sensors. For example, a GazeTag attached to a plant pot could report its soil moisture level (Figure \ref{fig:teaser}), allowing an AI assistant to proactively alert the user when it has been too long between waterings. 
Similarly, a GazeTag on a kettle could report its current temperature, allowing an AI assistant to gauge time to completion or alert the user when the water has reached an ideal temperature for tea vs. coffee (Figure \ref{fig:example-apps}A). 

\subsection{Personalization \& Access Control}
GazeTags can also tailor their response based on who is looking at them (user-worn devices transmit user credentials). One option is for GazeTags to only respond to their owner. Alternatively, GazeTags could reply with different information to different sets of users. As one example, a GazeTag on a fire extinguisher could report its object type to anyone, but include its last service date when seen by health and safety personnel (Figure~\ref{fig:example-apps}B). 

\subsection{Object-Integrated GazeTags with Actuation}
If integrated into objects, GazeTags could allow for actuation. For example, after reading all GazeTags in a scene, a user-device could follow up with a command, either triggered explicitly by a user or by an AI agent on behalf of the user. As a simple example, a user could look at a lamp and say ``turn off.'' The user-worn device would transmit the recipient object's UID and command, and the GazeTag attached to the lamp would toggle it off (Figure \ref{fig:example-apps}C). 

\section{Related Work}
In this section, we review related work on sensing user context. We organize by approach, first reviewing work that primarily uses computer vision, then systems that instrument the environment with passive tags, and conclude with active tagging approaches, which are most similar to GazeTags. 

\subsection{Sensing Context from Vision}
Cameras are perhaps the most common sensing modality for understanding user context, being widely used in both research~\cite{sigurdsson_what_2017, liu_no_2021, cv_context_survey} and commercial systems~\cite{noauthor_chatgpt_2026, upgrade_gemini_2026}. They can infer context from a user's gaze direction~\cite{konrad_gazegpt_2024, hansen_eye_2010,krafka_eye_2016, cheng_appearance-based_2024, recasens_where_2015, kim_hifigaze_2026}, their body and hand pose~\cite{pavlakos_reconstructing_2024, potamias_wilor_2025, goel_humans_2023} and any objects attended~\cite{redmon_you_2016, carion_end--end_2020, carion_sam_2026}. Pointing~\cite{fashimpaur_squiggle_2025, lee_gazepointar_2024, kim_worldpoint_2023, bolt_put-that-there_1980} and head-gaze~\cite{mayer_enhancing_2020} can further inform where in the world a user is referring to. More recently, cameras have also been used to infer user intent~\cite{handsonvlm, sigurdsson_what_2017, liu_no_2021, kovashka_learning_2010, feichtenhofer_slowfast_2019}, e.g., where are the user's hands moving next and what are they going to grab? While powerful, these approaches are power hungry, computationally expensive and privacy invasive, limiting their use on a battery-constrained, always-on wearable.

\subsection{Sensing Context with Passive Tags}
Another approach for sensing context is to instrument the environment with tags, read by a user-worn device. Tags can offer the benefit of increased identification accuracy, data-bearing payloads, and the ability to resolve visually identical objects (by using different IDs). We first discuss passive tags, which are either unpowered or receiving power from a user-worn transmitter.

The most pervasive passive tagging technology is barcodes, which can be read by a camera or close-range optical scanner. Later standards, such as AprilTags~\cite{olson_apriltag_2011}, ArUco~\cite{garrido-jurado_automatic_2014}, and ARTags~\cite{fiala_artag_2005} can reveal not only identity, but also 6DoF pose, allowing for spatial interactions. However, similar to other vision-based approaches, they require a camera and computer vision pipeline, which can be power hungry and raise privacy concerns. Moreover, to work at longer ranges, tags have to be unreasonably large (e.g., >10x10cm) so as to be visually obtrusive, or require very high resolution cameras. 

Another popular approach is to use passive RFID tags, which can communicate through RF backscatter~\cite{hutchison_wirelessly-powered_2006, buettner_rfid_2008, sample_design_2008} and have been used to detect activity context~\cite{smith_rfid-based_2005, li_idsense_2015, buettner_recognizing_2009, bhattacharyya_low-cost_2010, li_paperid_2016}. Unfortunately, RFID readers (which transmit) are generally large, activate a wide area, are limited in range to a few meters, and are energy expensive. For these reasons, they have not been integrated into wearables, and the range precludes them from the types of use we envision for GazeTags. NFC technology \cite{dierk_alterwear_2018, hutchison_wirelessly-powered_2006, ye_body-centric_2022} is highly related, and small, low-powered readers have been integrated into wearables. However, as the name suggests, NFC tags are near field, operating at centimeter or body scales, and thus would not work for the types of interactions we describe. 

\subsection{Sensing Context with Active Tags}
It is also possible to instrument the environment with active tags for context sensing. Popular wireless tag protocols include Bluetooth ~\cite{noauthor_tile_2026, ayyalasomayajula_bloc_2018}, ZigBee~\cite{gill_zigbee-based_2009, bianchi_rssi-based_2019}, and Wi-Fi~\cite{bahl_radar_2000, youssef_horus_2005}. Approximate distance to these tags can be estimated via RSSI and similar signals. Other approaches have been explored in the literature, including powered tags that communicate wirelessly through modulated light~\cite{want_active_1992, ahuja_lightanchors_2019, yang_infoled_2019, pathak_visible_2015, noauthor_optitrack_2026}, illuminated fiducial markers \cite{mohan_bokode_2009}, and  ultrasound~\cite{priyantha_cricket_2000, ward_new_1997}. In general, these are very energetic approaches that are not compatible with long term deployment on small batteries. To avoid this significant battery issue, researchers have also explored tags that generate their own energy through object actuation and interaction~\cite{zhang_sozu_2019, yang_cubesense_2023, yang_minikers_2022, fu_soundoff_2025}.

Among all prior work, the most closely related to GazeTags in terms of capability are UWB (ultra-wideband) tags, which exist as commercial products (e.g., Apple AirTags, Samsung SmartTags). These tags could, in principle, support head-gaze-aware interaction at room-scale ranges and can be made relatively low power~\cite{noauthor_airtag_nodate, zhao_uloc_2021, noauthor_estimote_2026}. The main downside is UWB remains expensive even after many years of commercial optimization. The bill of materials for an Apple AirTag is estimated at roughly \$10 (with a retail cost of \$29)~\cite{noauthor_apple_nodate}; the AirTag's Nordic SoC alone costs approximately \$4, even at large volumes \cite{noauthor_nrf52840-qiaa-r_nodate}. At 5$\times$ the cost of our GazeTags, dense deployments of hundreds of UWB tags across e.g., a household is not yet consumer feasible. Secondarily, although UWB needs only compact simple antennas for range-finding (as it relies on time of flight, not antenna directivity), multi-element array antennas are needed to estimate angle-of-arrival (AoA) on the reader device. Such antennas have yet to be demonstrated in small worn devices, such as earbuds or glasses (at present, the smallest UWB AoA implementation is in smartphones). Finally, UWB tags, like other RF approaches, can communicate without line of sight, and tags can be behind walls, inside cabinets, etc. This behavior is desirable when trying to find e.g., the key chain you lost in your house. However, it is undesirable behavior for our use case, where we wish to only know objects that lie in a user's field of view to contextually assist them.

\begin{figure}
    \centering
    \includegraphics[width=\columnwidth]{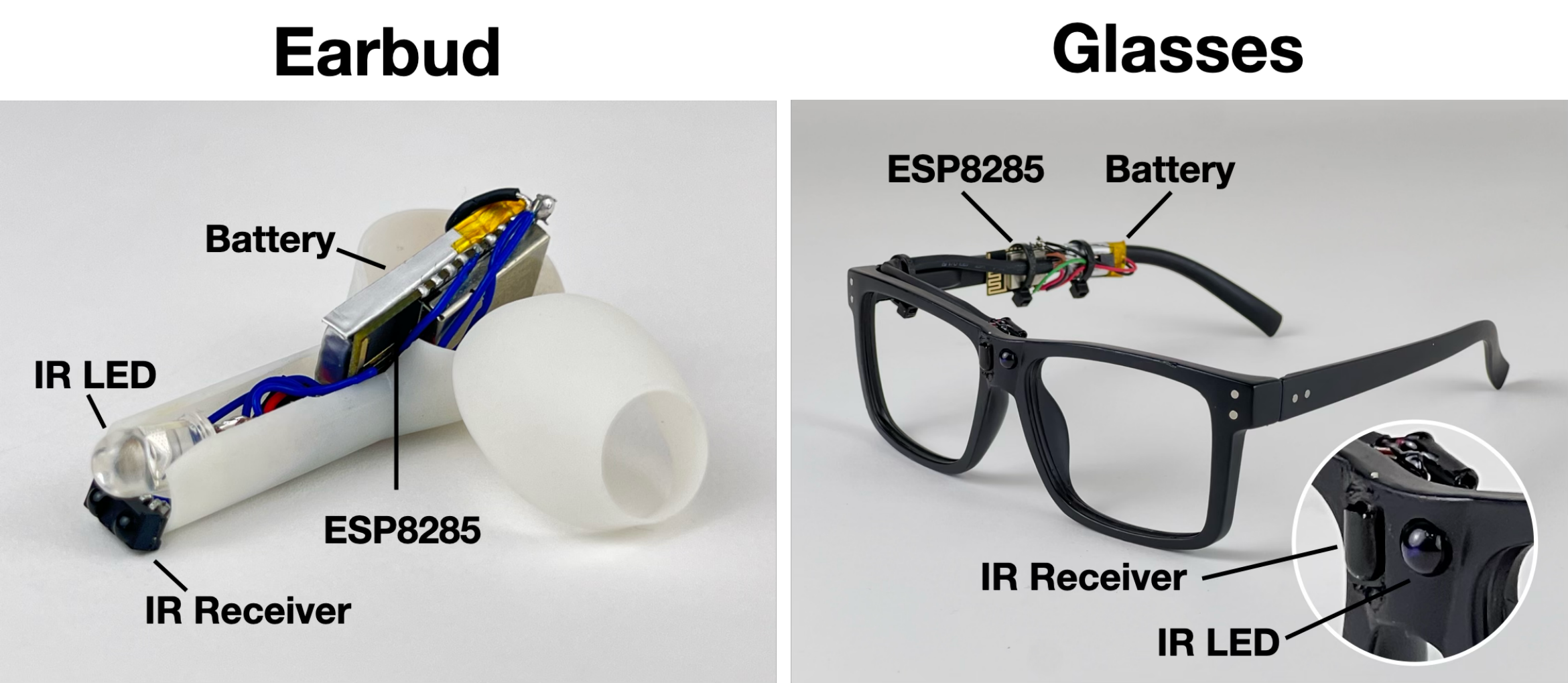}
    \caption{We created two head-worn prototypes to demonstrate the feasibility of GazeTags. }
    \label{fig:head-worn device prototypes}
\end{figure}

\subsection{Infrared Communication}
GazeTags utilize infrared (IR) communication, which has a long history as a simple, low-cost, power-efficient, wireless channel for short-range digital communication. In consumer electronics, IR has been used for many decades in remote controls for televisions, stereos, air conditioners, and related appliances. Beyond one-way command input, infrared also supported richer bidirectional device-to-device communication through the Infrared Data Association (IrDA) standards \cite{irda2001irphy14,microsoft2018irda}, which emerged in the mid-1990s and were widely adopted in laptops, PDAs, phones, printers, and cameras for short-range data exchange. 

Rather than introducing a new wireless substrate, GazeTags leverage existing infrared communication components. This brings several practical advantages: IR components are extremely low cost, highly available, and robust under everyday operating conditions, including bright light. As a result, GazeTags can inherit the reliability and affordability of an established technological ecosystem, repurposing it for gaze-aware object discovery.

We note that IR has previously been explored for identifying and selecting physical devices. Early among these is \textit{That one there!}~\cite{swindells2002that}, which used an IR-emitting stylus to point at devices, enabling two-way identification and subsequent wireless connection establishment. More closely related, HOBS~\cite{zhang2014hobs} used a glasses-worn IR transmitter to select IR-tagged objects based on head orientation. Like GazeTags, HOBS exploits the directionality of IR to associate a wearer’s head orientation with physical objects in the environment, and further explored techniques for refining selection when multiple tags fall within the IR field of view. Also very similar, InSight~\cite{boldu_insight_2017} used a glasses-worn IR transmitter (G-Ripple) and object-attached IR receiver tags (G-Sens) to pair arbitrary input controllers to smart devices such that the user could control objects they were facing. Each G-Sens "tag" device lasted only 50 hours on a single charge. We do not claim novelty in using head-directed IR for tagged-object selection. Instead, GazeTags explores a different point in this systems design space: whether such an approach can support dense, persistent instrumentation using tags that are ultra-low-cost, long-lived, small, and capable of reliably returning both identity and dynamic object state. Moreover, rather than focusing primarily on explicit object selection, we envision GazeTags continuously and passively exposing contextual object information to worn AI agents, which can use these signals to infer context and proactively support users’ ongoing tasks.

\begin{table}[]
\centering
\setlength{\tabcolsep}{2pt}
\renewcommand{\arraystretch}{0.8}
\begin{tabularx}{\columnwidth}{Xr}
\toprule
\textbf{Part}                 & \textbf{Subtotal}                  \\ \midrule
Microcontroller (ATtiny85)   & \$0.41                              \\
Infrared Receiver (TSOP18338) & \$0.41                              \\
MOSFET (2N7000, QTY: 2)       & \$0.08                              \\
Misc. Passives                      & \$0.04                              \\
Infrared LED (140$^\circ$ FoV)    & \$0.29                              \\
Battery (CR2477, 1100 mAh)    & \$0.60                              \\
Battery Holder                & \$0.26                              \\
PCB                           & \$0.06                              \\
Enclosure                           & \$0.05                              \\
\midrule
\textbf{Total}                & \multicolumn{1}{r}{\textbf{\$2.20}} \\
\bottomrule
\end{tabularx}
\vspace{5pt}
\caption{Bill of materials for a GazeTag. When applicable, subtotals are based on purchase volumes of 10K units.}
\label{tab:gazetag-bom}
\end{table}

\section{Implementation}
We now describe the hardware and software implementation of GazeTags and the user-worn devices that communicate with them.

\subsection{Head-Worn Device Prototype}\label{sec:head-worn device}

To demonstrate that GazeTags can work with a variety of head-worn form factors, we built two prototype devices: smart glasses and earbuds (Figure~\ref{fig:head-worn device prototypes}). Both use the same core hardware --- an ESP8285 SoC \cite{esp8285}, a TSOP18338 infrared receiver \cite{tsop18338}, and a 940~nm infrared LED. The latter wavelength was purposely selected as it is completely outside the visible spectrum (unlike common 790 or 850~nm infrared LEDs). Also, like previous infrared communication systems, we used modulated carriers (38 kHz) which offer robust operation under ambient light. We discuss selection of LED beam angle in the next section. The ESP8285 SoC was selected for its small footprint ($20.0 \times 12.3 \times 3.0$ mm) and built-in wireless connectivity.

Unlike our GazeTags, which must operate for long durations without maintenance to be practical, worn devices are typically charged daily. This significantly relaxes their power constraints, and thus our head-worn devices take on the more energy-expensive role of transmitting a wake-up signal. Specifically, to wake GazeTags, a head-worn device emits a 300~ms wake-up signal consisting of a 300~$\mu$s 38~kHz square wave (50\% duty cycle) followed by a 700~$\mu$s blanking interval (Figure~\ref{fig:timing-diagram}A). It then transmits a GazeTag signal with optional user credentials and waits for a reply from nearby tags. This transmission consumes 0.007~mWh of a head-worn device's battery (this is total energy usage, including the LED, ESP8285, and IR receiver). For reference, Meta Ray-Ban Display glasses contain a 960 mWh battery \cite{meta_rayban_display}. Of course, this wake-up signal would only trigger periodically, e.g., every 3~seconds, or be intermittently triggered by an AI assistant when it determines a user might benefit from contextual assistance. 

\subsection{Head-Worn Narrow Beam Transmitter}\label{sec:head-worn device narrow beam}
A tradeoff we make being camera-free is that we cannot track eye gaze, and must fall back to a less accurate gaze proxy: head direction. Fortunately, prior work has shown that users' eyes most often stay within a comfortable range of ±25º \cite{stahl_amplitude_1999}, and a majority of the time, eyes deviate from the head vector less than ±20º \cite{freedman_coordination_2008}. This informed the selection of our head-worn infrared transmitter: We selected an LED with a ±20º emission pattern. As LEDs are almost point light sources, the beam shape is roughly that of a cone.

\subsection{GazeTag Hardware}\label{sec:gazetag-hardware}

\begin{figure}
    \centering
    \includegraphics[width=\columnwidth]{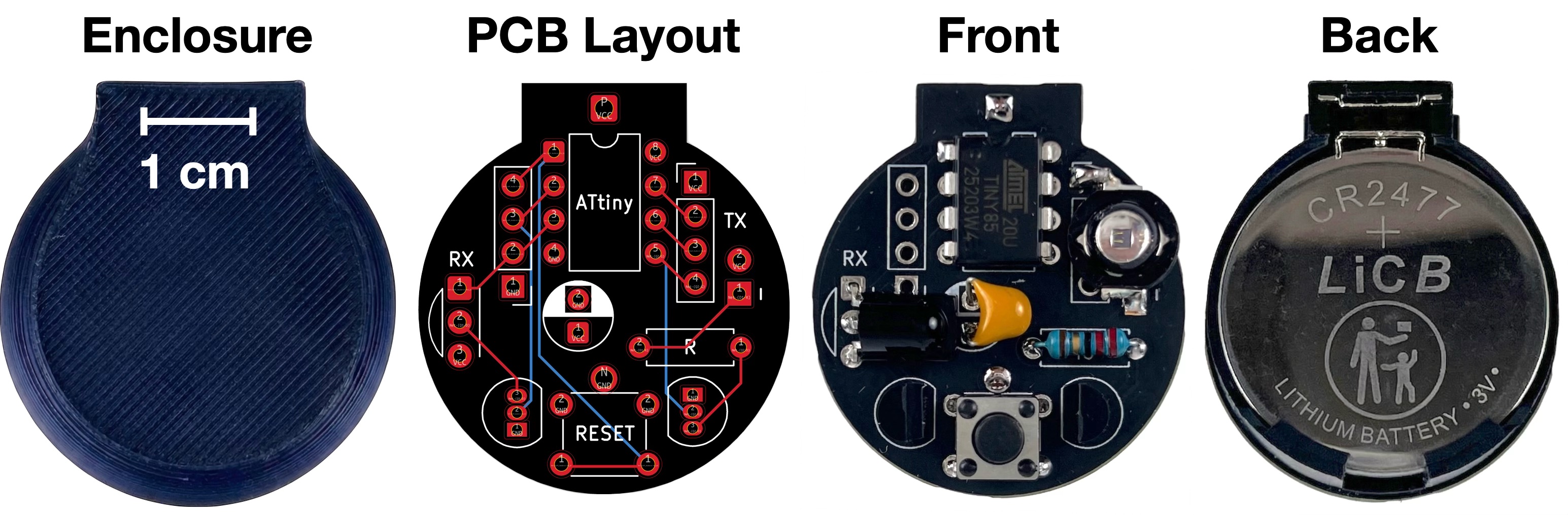}
    \caption{Different views of a GazeTag.}
    \label{fig:GazeTag-hardware}
\end{figure}

We designed our GazeTags (Figure~\ref{fig:GazeTag-hardware}) around the CR2477 coin cell battery, which measures 24.5~mm in diameter and 7.7~mm in height. The CR2477 provides 1100~mAh of capacity, which provides more than a year of runtime (discussed in Section~\ref{sec:battery-life}). We fabricated a custom PCB sized to match the CR2477's footprint (Figure \ref{fig:GazeTag-hardware}). The board hosts an ATtiny85 microcontroller, which we selected for its excellent deep sleep modes and low cost. The ATtiny85 drives a 940~nm infrared LED using a 2N7000 MOSFET. Unlike the narrower, cone-shaped beam LED used in the headset (Section~\ref{sec:head-worn device narrow beam}), the GazeTag uses a 140$^\circ$ conical emission LED. This wide emission pattern provides a large effective viewing range, regardless of how the tag is oriented on a surface or object. The tag uses the same TSOP18338 infrared receiver as the headset \cite{tsop18338}, but switches it on and off to conserve power, using another 2N7000 MOSFET (discussed in Section~\ref{sec:battery-life}). The front is enclosed in a friction-fit, 3D-printed case using an infrared-transparent filament~\cite{noauthor_pla_nodate}, allowing the LED and receiver to operate without cutouts. In total, each GazeTag costs \$2.20; a bill of materials is provided in Table~\ref{tab:gazetag-bom}.

\subsection{Communication \& GazeTag Firmware}
\label{sec:communication}

\begin{figure}[b]
    \centering
    \includegraphics[width=\linewidth]{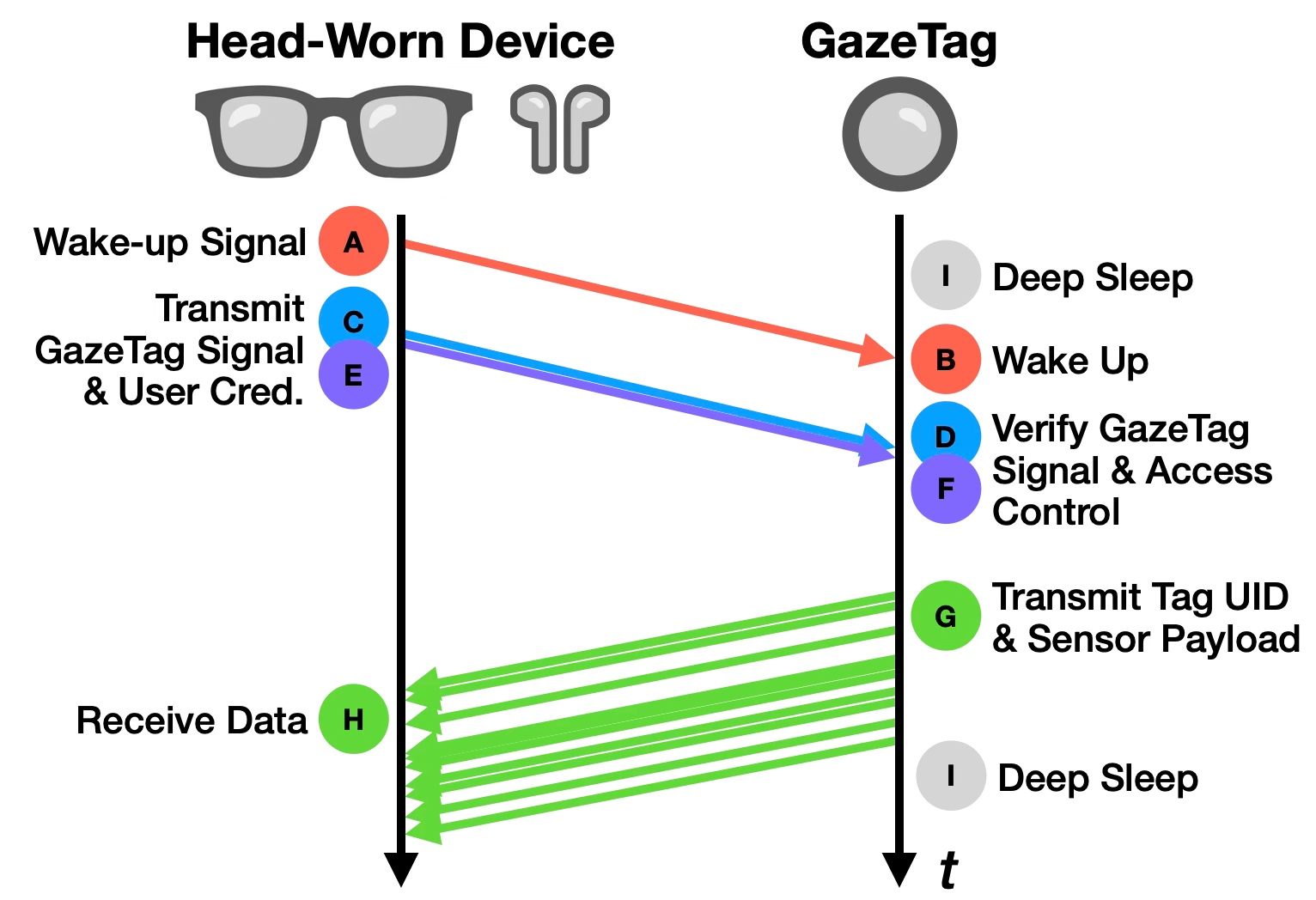}
    \caption{Overview of the communication sequence between a head-worn device and a GazeTag. Time not to scale.}
    \label{fig:timing-diagram}
\end{figure}

In this section, we provide an overview of how GazeTags and the aforementioned user-worn device communicate. As we step through this process, we also explain relevant GazeTag implementation details. Power consumption is discussed in Section \ref{sec:battery-life}. 

\subsubsection{Wake-up Interrupt}
GazeTags wake up from deep sleep (Figure~\ref{fig:timing-diagram}I) every 130 ms for 7~ms (i.e., 5.1\% duty cycle). During this brief interval, the GazeTag powers on its IR receiver. For the first 1 millisecond, the IR receiver is allowed to stabilize its internal automatic gain control (AGC). We then enable microcontroller interrupts on the IR receiver's digital pin and wait 6~ms. If no IR transmission is detected during this period, interrupts are disabled, the IR receiver is powered down, and the microcontroller goes back to deep sleep (Figure~\ref{fig:timing-diagram}I). If an IR transmission is detected, triggering an interrupt (Figure~\ref{fig:timing-diagram}B), we wait 350 ms to ensure the head-worn device has completed its wake-up signal (Figure~\ref{fig:timing-diagram}A) and finished transmitting any data (Figure~\ref{fig:timing-diagram}C/E). During this interval, interrupt handlers are used to buffer incoming IR data. Each time data is appended to the buffer, it is checked to see if it contains a valid transmission. If one is detected and correctly decoded, we set a flag for later processing. For all of our communication, we use the FAST IR protocol \cite{ir_FAST_protocol_github}.

\subsubsection{Head-Worn Device Sender Check}
In the IR receiver interrupt handler, GazeTags check for a valid sending device signal (Figure~\ref{fig:timing-diagram}D). This is a single FAST packet containing a fixed 8 bit ID --- we selected 0x68 ('h' for head device). This is necessary so that GazeTags do not reply to other IR transmissions in the environment, such as TV remote controls. Not only would this waste valuable battery power, but it would also interfere with the operation of other devices. If the IR transmission was not from an applicable GazeTag-capable device, the tag simply returns to deep sleep (Figure~\ref{fig:timing-diagram}I). 

\subsubsection{Access Control (Optional)}
The head-worn device can also transmit a user ID (Figure~\ref{fig:timing-diagram}E), allowing only authorized users to trigger a GazeTag response. Our current prototype uses a hard-coded UID as a simple access-control mechanism (Figure~\ref{fig:timing-diagram}F). This is sufficient for proof-of-concept use, but it is not secure in practice: a nearby adversary could observe the UID and replay it to impersonate the owner. A full implementation would use a rolling credential or lightweight cryptographic challenge-response protocol, preventing naive sniff-and-replay attacks. 

\subsubsection{Reply Payload}
The default reply of a GazeTag is a 16 bit object UID contained in two back-to-back packets, each taking 29 ms to transmit \cite{ir_FAST_protocol_github}. A user's head worn device could maintain their own UID-to-device lookup tables, or this could be a cloud service. A range of UIDs could be reserved to identify common objects (e.g., 0x4C9E = any water fountains), while a separate range could be reserved for personalized one-to-one mappings  (e.g., 0xA7F3 = water fountain, Newell Simon Hall, 1st floor). 

\subsubsection{Sensor Data (Optional)}
The microcontroller we use for our GazeTags has two unused pins that we expose as through-hole pads on our PCB. Both pins are capable of reading analog values, or interfacing with digital sensors. We describe some example uses for sensor data in Section~\ref{sec:example-apps}. When we want to reply with sensor data, we append an extra FAST packet to our reply payload. 

\begin{figure}[b]
    \centering
    \includegraphics[width=\linewidth]{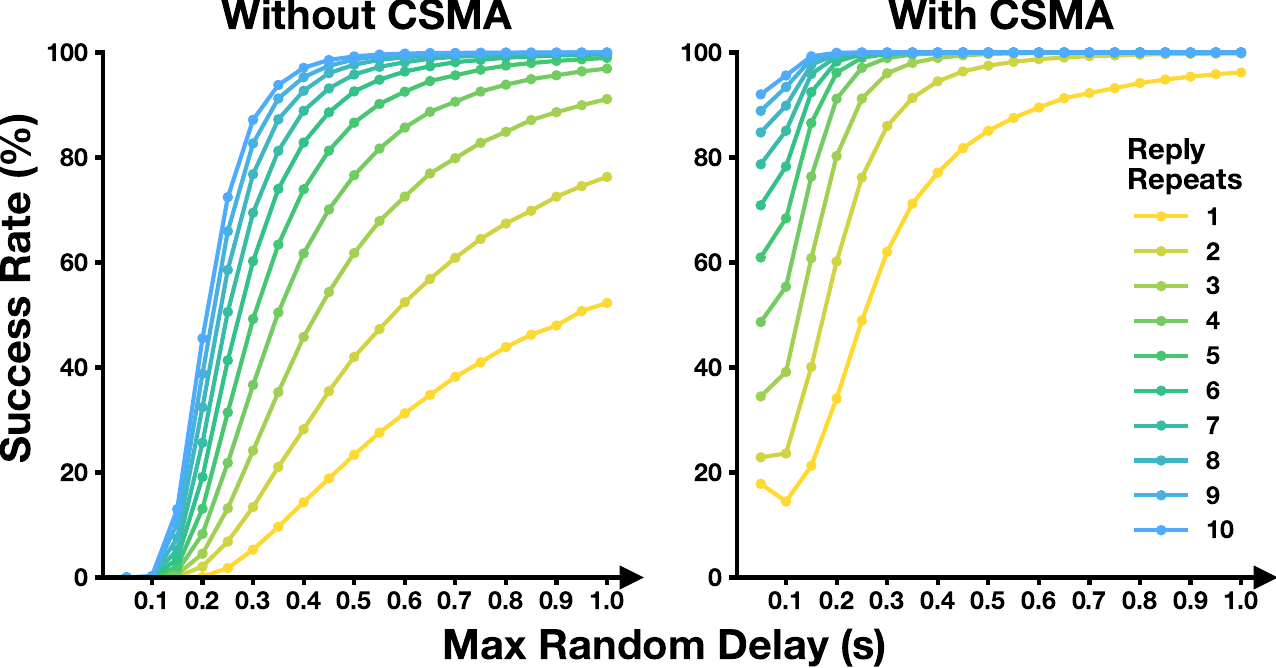}
    \caption{To inform our GazeTag collision avoidance implementation, we ran a Monte Carlo simulation with five co-located GazeTags. We vary the number of replies, delays between replies, and with or without CSMA. }
    \label{fig:csma-sim}
\end{figure}

\begin{figure*}
    \centering
    \includegraphics[width=\linewidth]{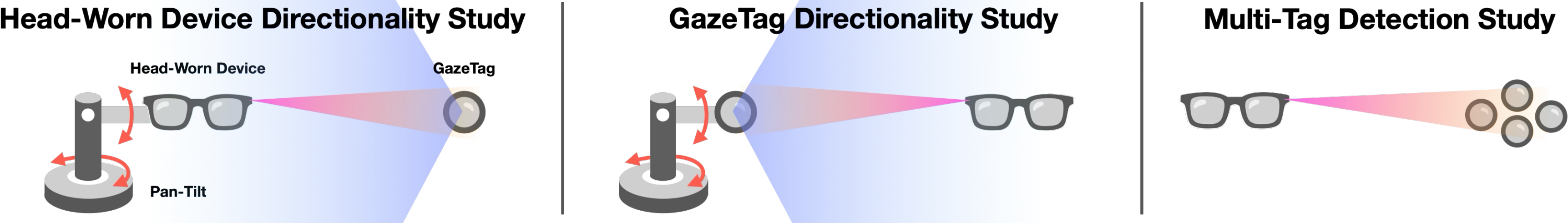}
    \caption{Illustration of the experimental setup for our three performance studies. Not to scale. Each study was repeated at 4, 6 and 8~m distances to also evaluate range. Results plotted in Figures \ref{fig:directionality-study} and \ref{fig:multitag-interference-study}. }
    \label{fig:robot-arm-setup}
\end{figure*}

\subsubsection{Multi-Tag Collision Avoidance}\label{sec:collision-avoidance}

If more than one GazeTag is in the receptive field of the head-worn device, multiple tags will compete to reply in a naive implementation, potentially leading to a subset or even no tags being recognized. To avoid collisions and ensure high detection reliability, we employ two schemes. First, GazeTags emit ten replies with random delays between 40 and 200 ms (Figure~\ref{fig:timing-diagram}G). Each 16-bit reply takes 58 ms to transmit, and so collisions are comparatively rare when there are just a few tags. Second, GazeTags implement carrier-sense multiple access (CSMA). Put simply, after waking from a random delay, GazeTags check to see if any other IR transmissions are currently active (i.e., carrier sense), and if so, go back to sleep anywhere from 30 to 50 ms to try transmitting later. This simple check significantly decreases collisions and increases reliability. 

To help inform the main variables of our collision avoidance implementation, namely number of reply repeats and random delay range, we created a Monte Carlo simulation tool. Plots of simulated data (with and without CSMA) are included in Figure~\ref{fig:csma-sim}. To achieve a $\ge$99\% success rate with 5 co-located tags, we selected 10 repeats, with a 200~ms max delay, with CSMA. We evaluate real-world performance with up to 6 GazeTags side-by-side in our later study (Section~\ref{sec:multi-tag detection study}).

\subsection{GazeTag Power Consumption \& Battery Life}\label{sec:battery-life}
We drew design inspiration from Apple AirTags, which last roughly one year with typical use and have a user-swappable coin cell battery. Similarly, we targeted one-year run-times with a replaceable coin cell battery. To achieve such long battery life, we had to significantly optimize our device's power consumption. For this, we made extensive use of a Nordic Power Profiler Kit II \cite{noauthor_power_nodate}.

One easy optimization was down-clocking the ATtiny to 1 MHz, which still offers plenty of compute for our needs. Our device spends \textasciitilde94.9\% of its time in deep sleep, during which it consumes just 4.7 $\mu$A. The IR receiver (TSOP18338) consumes 0.67 mA when continuously powered, and for this reason, we only wake to power it on intermittently to check for incoming IR transmissions. This accounts for 5.1\% of device time, during which the tag consumes 1.6 mA. Amortized, this means we have a mean continuous power draw of 86.1 $\mu$A. We use a CR2477 coin cell battery with a rated capacity of 1100 mAh and a nominal voltage of 3.0 V. This means our GazeTags have a maximum operational life of \textasciitilde1.4 years.

Of course, power is also consumed when the tag has to reply by turning on its IR LEDs, which is a comparatively rare, but power-expensive operation. The LED, when consistently on, consumes 74.7~mA. FAST packets are comprised of a 3.2~ms header, 25.2 ms payload, and 0.5 ms terminating mark \cite{ir_FAST_protocol_github} (29 ms total). IR transmissions are 50\% duty cycle, 38~kHz PWM signals, and on average 50\% of the payload bits are zeros (LED off, consuming no power), so the effective power draw is closer to 22.3 mA during transmission. 

As already described above, to ensure reliable transmission even in the presence of other tags and IR systems, GazeTags transmit ten replies (580 ms total transmit time). Thus a single full reply consumes 0.0062 mAh (all components combined). If we assume an average usage of twenty interactions every day for a year, this would consume 4.1\% of the total capacity of the battery (1100 mAh).

\section{Evaluation}
To evaluate the performance of GazeTags, we conducted three experiments examining the head-worn device's directionality, GazeTags directionality, and multi-tag detection. Figure \ref{fig:robot-arm-setup} provides an illustration of the three experimental setups. In all three experiments, we evaluate end-to-end accuracy (i.e., round-trip communication performance) across three challenging long ranges: 4, 6, and 8 m. 
\subsection{Head-Worn Device Directionality \& Range}\label{sec:transmitter-directivity}

As already discussed, for GazeTags to serve as a reliable proxy for head gaze, the head-worn infrared transmitter must have a sufficiently narrow and directional beam. To systematically characterize the transmitter's directionality, we conducted an automated experiment. Specifically, we mounted our head-worn transmitter on an SO-101 robot arm~\cite{so_arm100_2024} at a height of 1~m (Figure~\ref{fig:robot-arm-setup}). A single GazeTag was placed on a tripod at the same height, directly facing the transmitter. We tested three distances between the transmitter and tag: 4~m, 6~m, and 8~m, chosen to represent typical interaction distances in indoor environments such as homes and offices. We intentionally excluded shorter distances (e.g., 1--2~m), as detection at close range is expected to be near-perfect, and we wanted to test the system at more challenging ranges.

We controlled the robot arm and head-worn device with a Python script. The robot arm swept through yaw and pitch angles ranging from $-80^\circ$ to $+80^\circ$ in $10^\circ$ increments. At each orientation, the head-worn device emitted a wake-up signal followed by a GazeTag signal and waited for a reply (3500~ms timeout). As described in Section~\ref{sec:communication}, each GazeTag responds with up to ten reply transmissions to help overcome collisions with other infrared transmitting devices, including potentially other GazeTags in the environment. A trial was counted as successful if at least one reply was received by the head-worn device within the window. This process was repeated 10 times at each pitch--yaw--distance combination, yielding $17 \times 17 \times 10 \times 3 = 8{,}670$ total trials. The study was conducted indoors in typical office lighting and took approximately 4 hours to complete per distance.

\subsection{GazeTag Directionality \& Range}
Users may approach a GazeTag from a wide range of angles, and thus, for reliable operation, GazeTags must operate with a wide field of view. This not only includes receiving the head-worn device's wake-up signal, but also for transmitting a reply back to the sender (enabled by the wide-angle LED described in Section~\ref{sec:gazetag-hardware}). To systematically characterize our tag's receptive field, we conducted a second automated experiment.

We used the same apparatus and procedure as described in Section~\ref{sec:transmitter-directivity}, but with the positions of the GazeTag and head-worn device swapped. Specifically, we mounted a GazeTag on the robot arm and placed the head-worn device on a tripod at the same height (1~m), directly facing the tag. As before, a Python script controlled the robot arm and head-worn device. The robot arm rotated the GazeTag through yaw and pitch angles from $-80^\circ$ to $+80^\circ$ in $10^\circ$ increments at distances of 4~m, 6~m, and 8~m. The communication protocol, success criterion, and number of repetitions were identical, again yielding $17 \times 17 \times 10 \times 3 = 8{,}670$ total trials. This study also took approximately 4 hours to complete per distance.

\subsection{Multi-Tag Detection Study}\label{sec:multi-tag detection study}
In real-world use, multiple GazeTags may be within the field of view of the head-worn device, especially at longer distances. To test the efficacy of our collision avoidance implementation (Section~\ref{sec:collision-avoidance}), we conducted a third experiment. For this, we placed the head-worn device on a tripod pointed directly at a 25~cm wide shelf at the same height (1~m). We evenly spaced between 2 and 6 GazeTags on the shelf, with each tag programmed to transmit its own unique UID. As already discussed in Section \ref{sec:collision-avoidance}, we optimized our design to support 5 simultaneous GazeTags. Thus, we selected one above this quantity to see how accuracy would degrade. As with our previous studies, we tested at distances of 4~m, 6~m, and 8~m. The head-worn device emitted a wake-up signal followed by a GazeTag signal, and waited up to 3.5 seconds for any replies. We counted a trial as successful if at least one reply was received from every tag present. Each combination of tag count and distance was repeated 100 times. The study took a total of roughly 3 hours to complete.

\section{Results}
\begin{figure}[b]
    \centering
    \includegraphics[width=\linewidth]{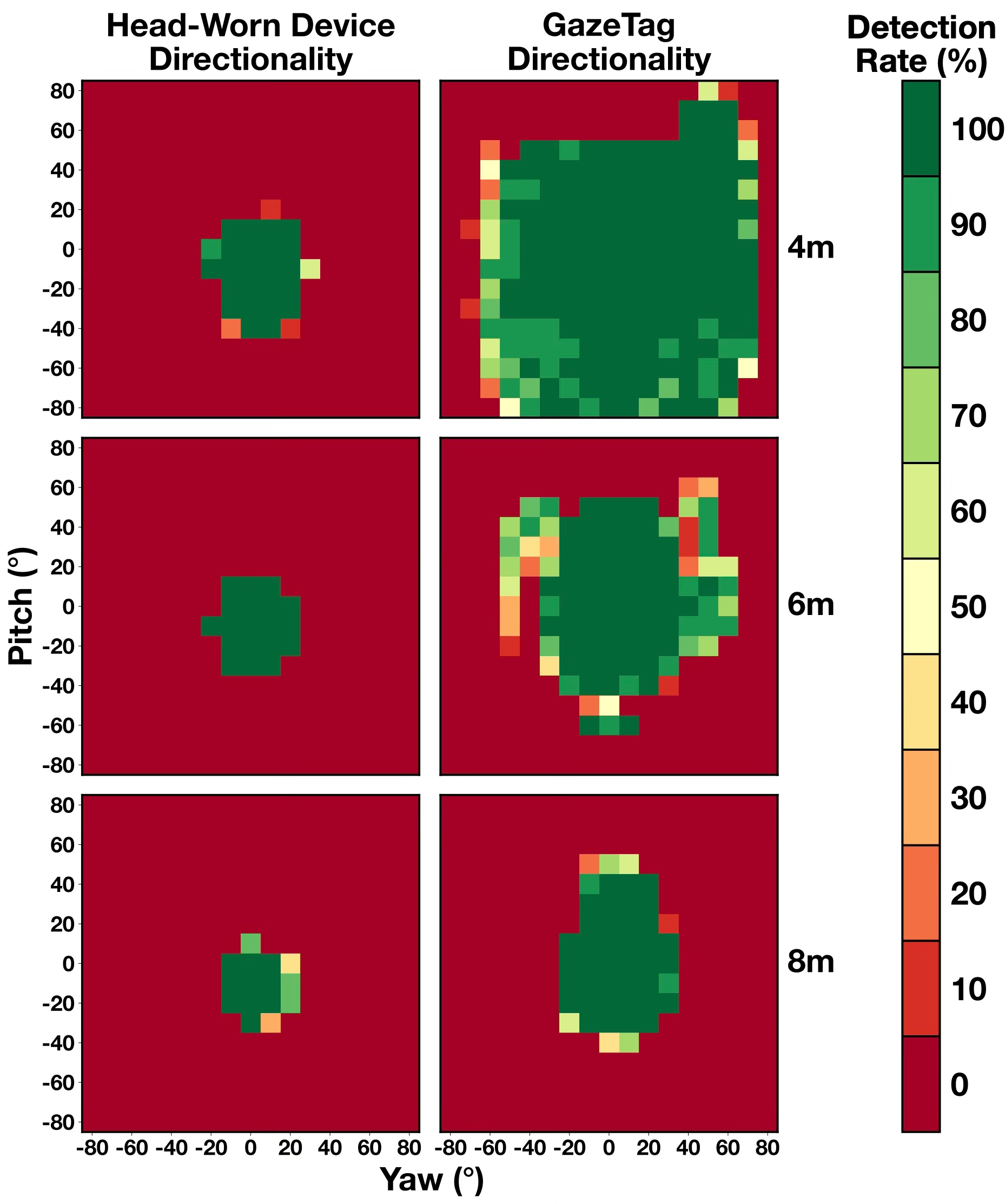}
    \caption{Results of our two directionality studies (see also Figure \ref{fig:robot-arm-setup}, left and center illustrations). Note the head-worn device has a purposely narrow beam shape, while GazeTags have a purposely wider one. }
    \label{fig:directionality-study}
\end{figure}

\begin{figure}[b]
    \centering
    \includegraphics[width=\linewidth]{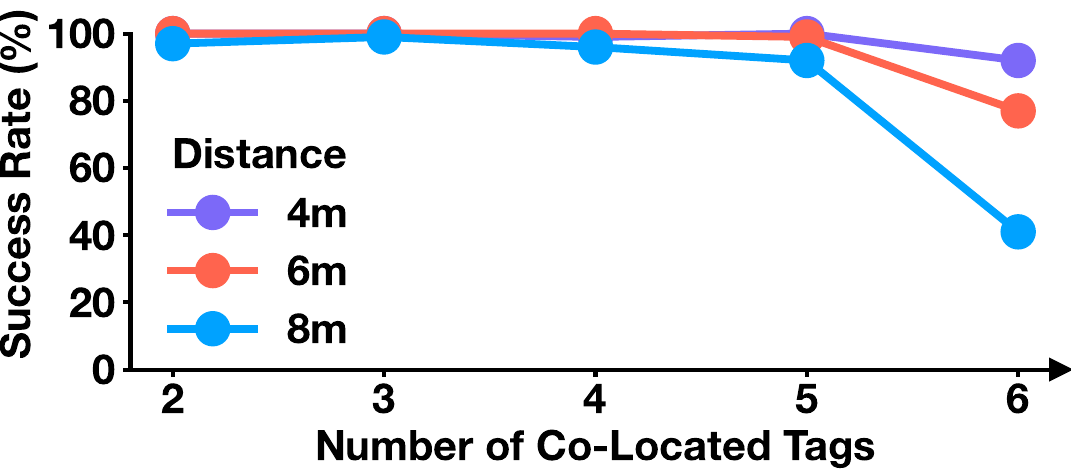}
    \caption{Results of our multi-tag detection study. GazeTag detection rates stay $\ge$99\% with up to 5 co-located tags and at distances up to 6~m.}
    \label{fig:multitag-interference-study}
\end{figure}

A summary of results can be found in Figures \ref{fig:directionality-study}--\ref{fig:multitag-interference-study}. Importantly, all accuracies reported in this section are computed as successful round-trip communications (i.e., head-worn device transmitting, GazeTag receiving, GazeTag transmitting, head-worn device receiving). At the high level, we found our system works well up to a range of 6~m, and operates in excess of 99\% reliably even with five GazeTags present within a small, 25~cm area.

\subsection{Head-Worn Device Directionality \& Range}
Results from our head-worn device directionality study are visualized in Figure~\ref{fig:directionality-study}, which characterizes the beam shape. Within the intended narrow beam, detection accuracy is near 100\%. There is a slight narrowing of the effective field of view as the range grows. This is because although the beam spreads linearly with distance, the light intensity falls off as the inverse square of distance. However, across all ranges tested, our head-worn device maintains a usefully narrow activation "spotlight" onto the scene. 

\subsection{GazeTag Directionality \& Range}
Results from our tag directionality study are visualized in Figure~\ref{fig:directionality-study}. At 4~m, the GazeTag exhibits an almost complete $180^\circ$ field of view with near-perfect detection rates. This means that users can approach a GazeTag from nearly any angle, including perpendicular to its normal axis (i.e., a user leaning against a wall, looking at a GazeTag attached to the same wall meters away). There is an obvious asymmetry in the accuracy on the vertical axis (Figure~\ref{fig:directionality-study}, right), which is due to the design of the IR receiver we used (which has a slightly asymmetric FoV \cite{tsop18338}). As with our previous study, the receptive FoV narrows as distance increases due to the drop-off in transmitted infrared intensity. 

\subsection{Multi-tag Detection}
Results from our multi-tag detection study are plotted in Figure~\ref{fig:multitag-interference-study}. With up to five co-located tags in a small, 25 cm area, performance is strong across all distances: the average detection rate is 99.75\% at 4~m, 99.75\% at 6~m, and 96\% at 8~m. With 6 tags, accuracy drops more significantly to 92\% at 4~m, 77\% at 6~m, and 41\% at 8~m. We optimized our system to reliably support up to five co-located tags (Section \ref{sec:collision-avoidance}), as we felt this represented a reasonable upper bound for the number of tagged objects a user would be facing in many settings. Supporting additional tags is possible by increasing the number of reply repeats and extending the random delay range.

\section{Limitations and Future Work}
Head gaze is inherently less fine-grained than eye gaze. While prior literature has shown that eye gaze typically deviates from head gaze by no more than approximately $20^\circ$~\cite{freedman_coordination_2008, adapting_coordination_eyes_head}, this is not always the case. If a user was to rotate only their eyes when glancing at nearby or peripheral objects, GazeTags would not work as reliably. For many use cases, however, we believe head-oriented gaze is a sufficient proxy for eye gaze, and the narrow beam of our head-worn device helps mitigate this issue.

GazeTags require line-of-sight between the user and the GazeTag in order to function, which could be obstructed through self-occlusion. In our glasses prototype, the hands could block line-of-sight when raised in front of the user, but this is less common when attending to objects meters away. We envision our IR hardware to exist on both earbuds, such that they have a complete FoV when operating together. However, earbuds would still have to contend with long hair, which is an innate limitation of this placement.

In our current implementation, the head-worn device emits a wake-up signal every three seconds, which offers a reasonable latency for many context-aware use cases. GazeTags can also operate in a more event-driven manner, where the head-worn device emits a wake-up signal in response to certain triggers. For instance, an AI assistant could decide to emit a wake-up signal when it needs additional context (e.g., which room is the user in? What objects are they facing?). 

As compared to a purely computer-vision-based approach, GazeTags require instrumenting objects in the environment, which could be costly and visually obtrusive. With our current prototype that uses off-the-shelf components, it would cost roughly \$220 to instrument 100 objects across a home. Using SMD components and a smaller microcontroller (e.g., ATtiny412) would significantly reduce size. If the cost could be reduced to under \$1 and the form factor shrunk to around 1~cm, GazeTags would begin to approach a vision of computing as an infrastructure medium --- small, inexpensive, and ubiquitous enough to be embedded into nearly any object~\cite{smartdust}.

Although power-optimized, GazeTags still require batteries, and the CR2477 coin cell will eventually need replacement. With further power optimizations, GazeTags could operate on a smaller coin cell or run off of harvested solar energy. A mains-powered version of a GazeTag could also enable longer-range ($\ge$20 m) use cases. For instance, buildings and shops could broadcast menus and other information as people look towards them. We note that consumer TV remote controls use the exact same infrared components and have been highly engineered to work outdoors at long range, so while we did not evaluate outdoor performance, we do not anticipate this to be a fundamental limitation.

Finally, head-worn devices could also act as GazeTags. For instance, at a conference, two attendees wearing GazeTag-equipped devices could exchange contact information simply by facing each other, enabling a natural and effortless form of social networking.

\section{Conclusion}
We presented GazeTags, an inexpensive, low-power, camera-free approach to head-oriented object detection. GazeTags cost roughly \$2 each and run for more than a year off a single coin cell battery. Our studies reveal that GazeTags can work well up to a range of 6~m, and with over 99\% reliability with up to five co-located tags. 

\begin{acks}
We thank Andy Kong for his early work on this project.
\end{acks}

\balance

\bibliographystyle{ACM-Reference-Format}
\bibliography{references}

\end{document}